\documentclass[conference]{IEEEtran}
\makeatletter
\def\ps@headings{%
\def\@oddhead{\mbox{}\scriptsize\rightmark \hfil \thepage}%
\def\@evenhead{\scriptsize\thepage \hfil \leftmark\mbox{}}%
\def\@oddfoot{}%
\def\@evenfoot{}}
\makeatother
\usepackage{array}

\usepackage{multirow}
\usepackage{longtable}

\usepackage{indentfirst}

\usepackage{url}
\usepackage{cite}

\ifCLASSINFOpdf
   \usepackage[pdftex]{graphicx}
   \usepackage{epstopdf}
   \usepackage{float}

\else

   \usepackage[dvips]{graphicx}

\fi

\usepackage{bbding}

\usepackage{algorithmic}
\usepackage{algorithm}

\usepackage{amsmath}
\usepackage{amssymb}

\usepackage{array}

\usepackage{bm}

\usepackage{comment}

\usepackage{listings}

\usepackage{enumitem}

\usepackage[tight,footnotesize]{subfigure}

\title{A Secure Proxy-based Access Control Scheme for Implantable Medical Devices}

\usepackage[bottom=1.05in,top=0.75in,left=0.635in,right=0.64in]{geometry}

 \author{
 \IEEEauthorblockN{Longfei Wu\textsuperscript{$^{1}$},  Haotian Chi\textsuperscript{$^{2}$}, Xiaojiang Du\textsuperscript{$^{2}$} }

 \IEEEauthorblockA{\textsuperscript{$^{1}$}Department of Mathematics and Computer Science, Fayetteville State University, Fayetteville, NC, USA, 28301\\}

\textsuperscript{$^{2}$}Department of Computer and Information Science, Temple University, Philadelphia, PA, USA, 19122\\

 \textsuperscript{} Email: {$^{1}$}lwu@uncfsu.edu, {$^{2}$}\{tug66074, dux\}@temple.edu\\  
 }

\begin{document}
\maketitle

\begin{abstract}

With the rapid development of health equipments, increasingly more patients have installed the implantable medical devices (IMD) in their bodies for diagnostic, monitoring, and therapeutic purposes. IMDs are extremely limited in computation power and battery capacity. Meanwhile, IMDs have to communicate with an external programmer device (i.e., IMD programmer) through the wireless channel, which put them under the risk of unauthorized access and malicious wireless attacks. In this paper, we propose a proxy-based fine-grained access control scheme for IMDs, which can prolong the IMD's lifetime by delegating the access control computations to the proxy device (e.g., smartphone). In our scheme, the proxy communicates with the IMD programmer through an audio cable, which is resistant to a number of wireless attacks. Additionally, we use the ciphertext-policy attribute-based encryption (CP-ABE) to enforce fine-grained access control. The proposed scheme is implemented on real emulator devices and evaluated through experimental tests. The experiments show that the proposed scheme is lightweight and effective.


\end{abstract}


\section{Introduction}

IMDs are the particular type of medical devices that are implanted in the patient's body, to diagnose, monitor, or treat a variety of conditions, diseases and injuries. 
For example, insulin pump can monitor and deliver insulin to treat diabetes, pacemaker regulate the beating of the heart using electrical impulses, neurostimulator sends electrical signals to the spine to treat chronic pains.
According to a recent report published by Allied Market Research \cite{market}, the global IMD market is projected to reach \$116.3 billion by 2022. 
However, IMDs are threatened by both external cyber attacks and internal flaws in software or firmware design. These security vulnerabilities allow an adversary to steal sensitive medical data, reset the configuration parameters, and issue unauthorized commands to an IMD, which could cause fatal consequences.

IMDs are equipped with a radio transceiver to communicate with the external IMD programmer. The IMD programmer is the specific device used to collect the medical data from IMDs and issue operation/configuration commands to deliver drug, change dosage, etc. With the wireless interface enabled, IMDs can be accessed by an authorized operator in physical proximity via the IMD programmer  (e.g., an eligible medical staff or the patient himself/herself).
%
However, the wireless communication and networking capabilities of IMDs turn out to be the major sources of security vulnerabilities. 
Due to the broadcast nature of wireless channels, all messages exchanged between the IMD and the programmer can be captured by eavesdroppers. This would not only expose the patient privacy like he/she is carrying an IMD to treat a certain disease, but also lead to other classic wireless attacks such as the forging, tampering, and replying of the messages.
Existing research works have presented the breaches in a number of commercial IMDs \cite{SP2008, HealthCom2011, hack2011, hack2016, Jack}, including the implantable cardioverter defibrillator (ICD), insulin pump and pacemaker. It has been demonstrated how an adversary can reverse-engineer the communication protocol and take full control of the IMD using a software radio.

Intuitively, to secure the communications between the IMD and the IMD programmer, a pair of symmetric keys must be shared between the two parties to encrypt their wireless communications. 
Unlike traditional electronic devices, the power supply of IMDs is highly constrained. The wireless charging technologies for IMDs still need lots of practical testing and clinical trials to ensure that no negative effect will be caused to human organs and tissues. Additionally, the replacement of an IMD or its battery requires invasive surgery. Hence, commercial IMD products are designed to last for 5 to 10 years. The energy consumption of IMDs should be minimized, by avoiding complicated cryptographic computations and long-range wireless communications.
Currently, only symmetric cryptography is considered for the data encryption in IMDs.

In this paper, we propose a novel proxy-based access control scheme for IMDs, in which the communications between the IMD programmer and the proxy are conducted through an audio cable rather than the conventional wireless channels. The proposed scheme employs the attribute-based access control model, which grants access based on the attributes (i.e., qualifications) that the access requestor owns. Meanwhile, the access requestor is authenticated in our scheme, mainly to provide accountability in case of a medical dispute.

Our major contributions can be summarized as follows:

\begin{enumerate}[leftmargin=*]
\item We first comprehensively studied and analyzed various existing IMD access control schemes in terms of the access control architecture and access control model. We also took a full consideration of the special use cases that a good IMD control scheme should be able to handle.
\item Our proxy-based IMD access control scheme can greatly alleviate the computational overhead and power consumption of IMDs. The proxy communicates with the IMD programmer via the audio cable, which can defend against the wireless passive and active attacks. Unlike USB connection, communications through headphone jack does not require the patient to unlock the device for manual approval, which is a practical concern especially for patients who are unconscious.
\item Our scheme adopts the ciphertext-policy attribute-based encryption (CP-ABE) to provide a fine-grained access control over the qualifications of the programmer operator. Specifically, the proxy encrypts the temporary session key with a specific access policy and sends the ciphertext to the IMD programmer. If the programmer operator owns the set of required attributes (qualifications), the temporary session key can be correctly retrieved. 
\item We implemented our scheme on real emulator devices: the IMD is emulated by TelosB mote, the proxy is emulated by smartphone, and the IMD programmer is emulated by Rasberry Pi 3. The evaluation results show that the proposed scheme is lightweight and effective. 
\end{enumerate}


\section{Background}

IMD manufacturers are supposed to give equal attentions to the security of their products, as to their functionalities. However, when facing security vulnerabilities, the manufacturers that should take full responsibility seem to be numb towards the potential security problems. 
In May 2014, an independent security researcher Billy Rios discovered 100 vulnerabilities in the communications system software of several different versions of infusion pumps made by the medical device company Hospira (HSP), which can be exploited by an attacker to hack into the pumps and change the dosage of medication to be delivered \cite{CNN}.
Rios immediately notified Hospira, but Hospira stayed silent on the issue until another researcher Jeremy Richards publicly disclosed the vulnerability in April 2015. Then, the U.S. Food and Drug Administration (FDA) and the Department of Homeland Security's Industrial Control Systems Cyber Emergency Response Team followed up the issue, and sent out advisories notifying hospitals of the danger of Hospira pumps and encouraging the transition to alternative infusion systems \cite{Hospira}.
Although the governmental agency FDA is obliged to supervise and regulate the IMD industry, it only provides guidelines and recommendations for IMD security which are not legally binding \cite{premarket, postmarket}. There is no checking or verification of new IMD products (software and hardware) and their cybersecurity documentations by a trusted agency.

The security and robustness of IMDs still rely on the research and development team of each individual manufacturers, who design access control schemes only specific to their own products.

\begin{figure*}[t]
\centerline{
\subfigure[Two-party Access Control]{\includegraphics[width=3in,height=1.4in]{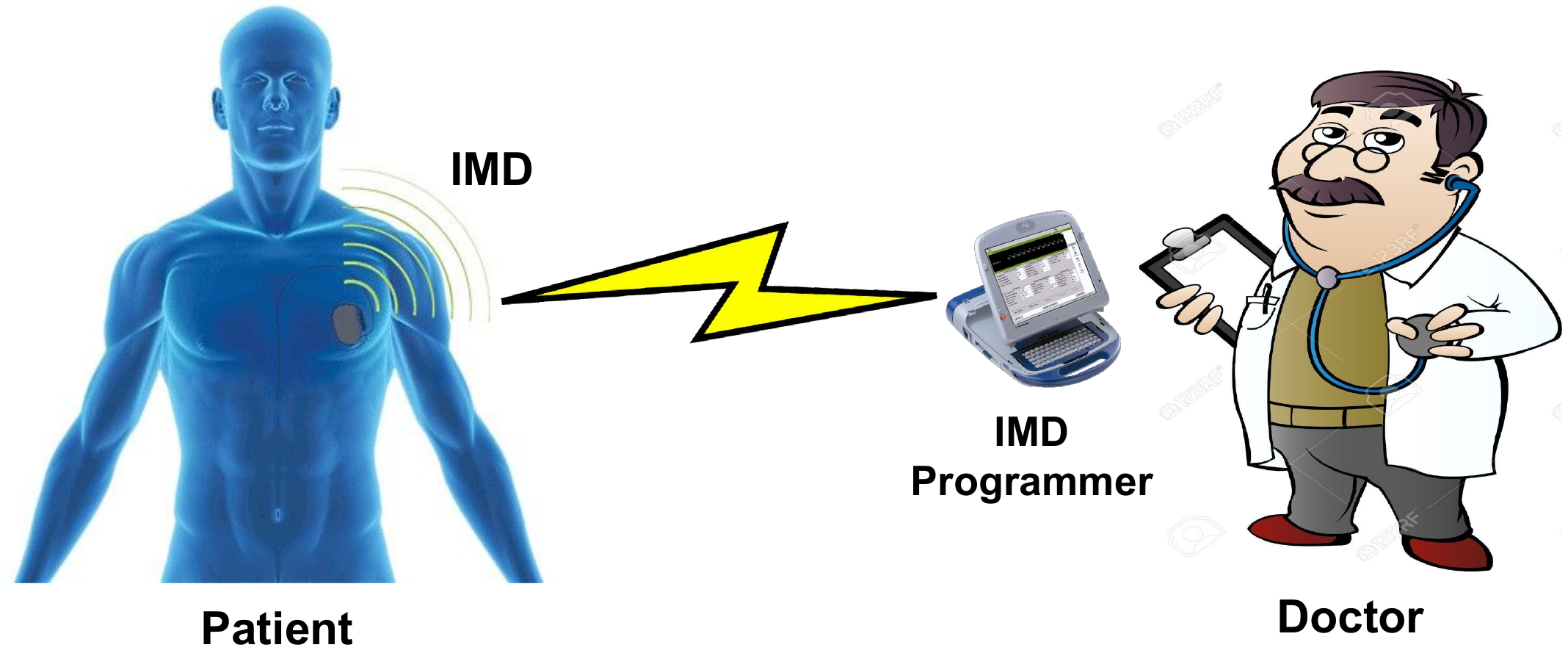}%
\label{fig:two}} %
\hspace{7mm}
\subfigure[Proxy-based Access Control]{\includegraphics[width=3.8in,height=1.5in]{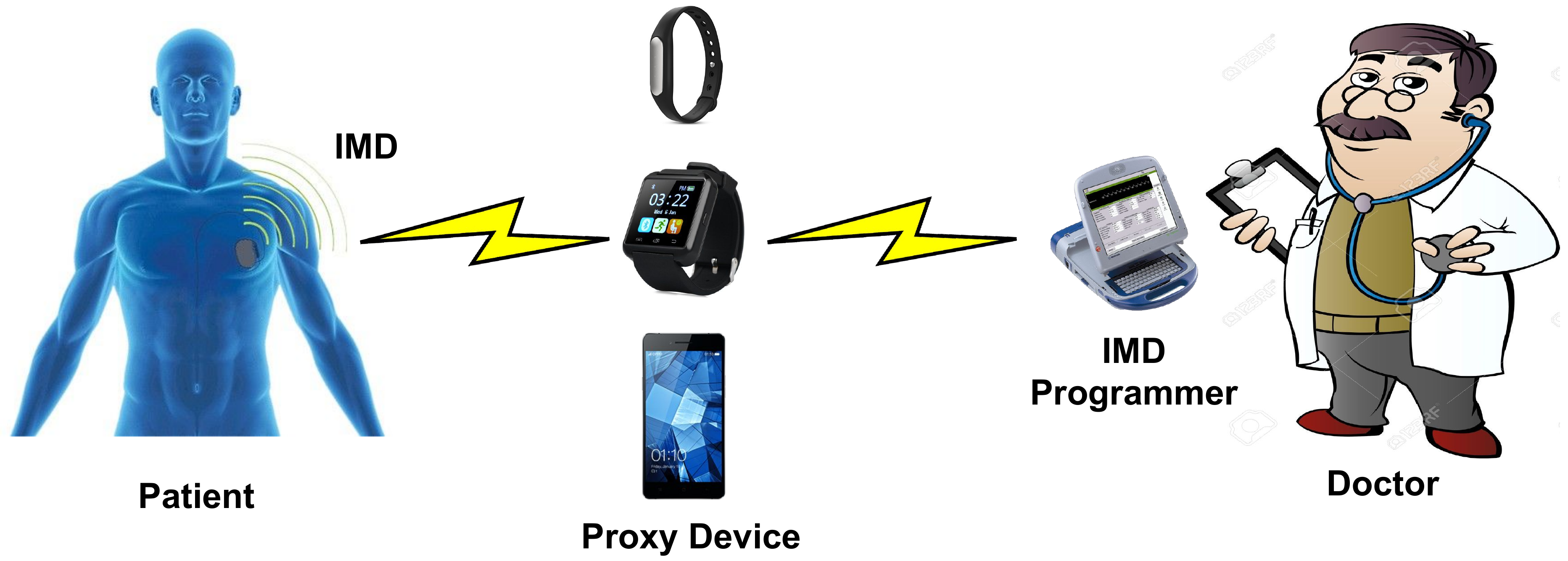}%
\label{fig:proxy}} %
}

\caption{IMD Access Control Architecture}
\label{fig:random}

\end{figure*}

\section{Motivation} 

The security researchers have been seeking generalized and effective access control schemes for IMDs. 
A variety of access control schemes have been proposed with different access control models and architectures. In addition, various assumptions have been made on the environmental settings and human factors. 
In this section, we conduct a thorough analysis on all these aspects, and present how we are motivated to design our novel IMD access control scheme.

%


\subsection{Special Use Cases}
IMD access control is not a difficult problem under regular situations such as when the patient uses his/her own IMD programmer to access the IMD or when an acquainted doctor wants to access the IMD. However, it becomes much more complicated in some special but practical situations.


\subsubsection{\textbf{Medical Emergency}} 
In medical emergency conditions such as when the patient falls sick while travelling out of town and needs immediate treatment, the patient has to verify the authenticity and qualification of the stranger who attempts to access the IMD (e.g., emergency medical technician). In a worse case, the patient may have been unconscious and is not able to manually verify the programmer operator. Hence, to deal with the emergency cases, an effective IMD access control scheme requires:

\begin{enumerate}[leftmargin=*]
\item All eligible medical personnels should have access to the IMD, regardless of whether they have been granted access before or if the patient is acquainted with them.
\item The access decision can be made autonomously by the IMD, without the patient's involvement. 
\item If the access is permitted, the IMD must first be paired up with the programmer so that a pair of symmetric keys are shared between them, to encrypt their communications. 
\end{enumerate}

\subsubsection{\textbf{Internet Connection}} 


Online authentication has been widely used in IT applications. Intuitively, offloading authentication to a dedicated server of a governmental health agency or hospital can greatly reduce the complexity of the access control computations running on IMDs \cite{online1, family}. This requires either the IMD (may be assisted by an external proxy) or the programmer to be able to connect to the Internet. However, Internet connection may not be always available, especially when the patient is located in a depopulated area with poor infrastructure. 
Hence, a robust IMD access control scheme should not rely on online authentication.


\subsection{Adversary Model}

The common assumptions agreed by existing works regarding the capabilities of the attackers in IMD context include:

\begin{itemize}[leftmargin=*]
\item The adversary may be equipped with powerful software radio transmitter, hence is able to interact with the IMD in a long distance. 
\item The adversary may obtain legitimate IMD programmer to access the IMD. Since IMD programmers are specialized devices running closed-source software programs, they are considered secure and cannot be hacked by the attacker. 
\item The adversary cannot approach the patient within a security range (typically 10 cm), nor can the adversary make physical contact with the patient or the patient's personal belongings, deterred from leaving criminal evidence such as fingerprint or video taken by surveillance cameras. 
\end{itemize}

Generally, two types of adversaries may exist depending on the attack tactics: passive adversary and active adversary. A passive adversary will only eavesdrop on the wireless channel and listen to the packets exchanged between the IMD and the IMD programmer. Given an unencrypted radio channel, a passive attack can break the confidentiality of the data being transmitted. Almost all existing access control schemes require encrytions over the wireless channel, hence are resistant to passive attackers. An active adversary, however, can replay or tamper the packets. If the communication protocol between the IMD and the programmer is reverse-engineered, the active attacker is able to send unauthorized commands to the IMD (e.g. changing the configurations and parameters).
Based on the purposes of the attacks launched by an active adversary, we can classify active attacks into three categories:

\begin{itemize}[leftmargin=*]
\item \textbf{Unauthorized access.} The goal of this type of attacks is to bypass the access control scheme and gain access to the IMD without authorization. 
\item \textbf{Resource depletion.} This type of attacks repeatedly requests for access to the IMD, causing the IMD to continuously running the access control computations, while its actual intention is to drain the battery power and reduce the lifetime of the IMD.
\item \textbf{Denial-of-service (DoS).} DoS attacks aim to disrupt the authorized access to the IMD, by interfering the communications between the IMD and the programmer. As the result, the IMD is unable to serve the incoming requests.
\end{itemize}

The resource depletion attacks and DoS attacks are not the focus of this paper. 
Many previous works have proposed effective schemes to solve the resource depletion attacks on IMDs. Liu \emph{et al.} \cite{Liu} suggested to add an extra wake-up circuit before the main circuit of the IMD, which employs the passive RFID technology so as to harvest energy from the incoming signal to perform the verification of the wake-up code. The main circuit is waked up only if the wake-up code is correct. Gollakota \emph{et al.} \cite{SIGCOMM} presented \emph{Shield} to protect the confidentiality of the IMD, which utilizes a novel full-duplex radio design with a jamming antenna and a receive antenna, allowing it to  simultaneously receive the IMD's signal and jam the IMD's messages. Consequently, the programmer cannot receive IMD's packets or directly interact with it. Hei \emph{et al.} \cite{Hei2010} proposed to train the normal IMD access patterns and detect unusual access requests using the support vector machine (SVM). 
The DoS attacks, however, have been evaded in existing works. Since the IMD is installed inside human body and has to interact with the programmer over a wireless channel, an attacker can just block/interfere its communications by jamming the wireless channel. Although the DoS attacks can be easily detected, there is no effective and low-cost solution to prevent it.

Instead, our paper focuses on enhancing the IMD access control in terms of lower complexity and better granularity.

\subsection{Access Control Architecture}

\subsubsection{\textbf{Two-party Access Control}} The basic IMD access control architecture is composed of two parties. As illustrated in Figure~\ref{fig:two}, the access object is the IMD and the subject is the programmer and its operator. 

\subsubsection{\textbf{Proxy-based Access Control}}
To reduce the energy consumption of the IMD, the proxy-based access control architecture has been proposed \cite{Hei2010,IMDGuard,Cloaker,MedMon}, which take advantage of a proxy device to delegate the heavy computations for the IMD. As shown in Figure~\ref{fig:proxy}, the proxy can be a smartphone or other wearable device (e.g., smart watch, smart bracelet) with more sufficient computational resources and battery capacity. 
%
%
%
The communications between the IMD and the proxy are protected by the lightweight symmetric encryption, which can be considered safe given that a pair of symmetric keys have been distributed and shared securely during the initial setup. This is reasonable since the initial setup is conducted either by the doctor when the IMD is implanted or by the patient when the proxy is used for the first time. It is very difficult for a malicious eavesdropper to overhear the key being transmitted to IMD at these specific moments. After pair-up, the proxy device will perform the access control on behalf of the IMD. 

The proxy-based access control depends on the presence of the proxy device. In the particular case that the proxy is not detected in vicinity, the commonly used solution is that the IMD will enter the open-access mode, in which it only verifies the physical proximity of the programmer and permits the incoming access requests from any programmer nearby. This allows eligible physicians to be still able to access the IMD when the proxy is damaged or lost. Later, patient can pair up the IMD with a new proxy using its unique master key. A copy of this master key is provided to the patient by the IMD manufacturer along with the product manual. Meanwhile, the IMD manufacturer or the hospital should keep a backup copy of the master key for the patient to retrieve from.

However, we are aware that the usage of a proxy device may increase the attack surface of the IMD access control. Two types of attacks targeted at the proxy may exist: 
\begin{itemize}[leftmargin=*]

\item \textbf{Jamming attacks.} The adversary may attempt to bypass the access control performed by the proxy, by selectively jamming the messages of the proxy to spoof the IMD about the absence of  the proxy. 
%
\item \textbf{Malware-based attacks.} If the proxy is a general-purpose device like smartphone, the adversary can attack through a pre-installed malware. However, since Android OS and iOS both use sandboxing to isolate applications from each other, the malware cannot compromise the client application running the access control program. Instead, it can only eavesdrop or disrupt the communications between the proxy and the IMD/programmer. 
%
%
\end{itemize}

\begin{figure*}[t]
\centering
\includegraphics[width=6.1in,height=1.3in]{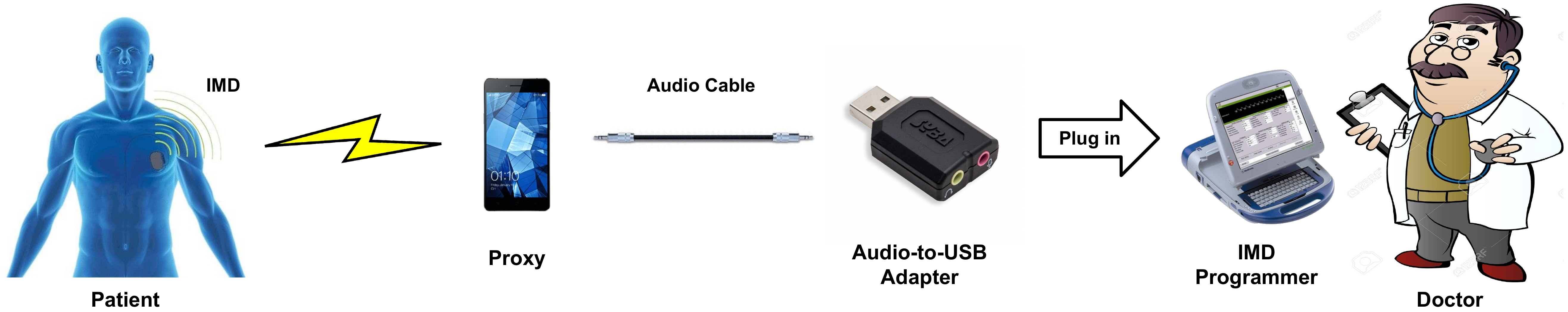}
\vspace{-2mm}
\caption{Architecture of Access Control with Wired Connection}
\label{fig:wired}
\vspace{-4.5mm}
\end{figure*}

Our scheme employs the proxy-based access control architecture. Note that we do not specifically address the jamming-based spoofing attack - our access control scheme can nicely integrate with existing solution \cite{IMDGuard}. Additionally, we do not propose new access control method for the particular case in which the proxy is indeed absent. The existing coarse-grained proximity-based access control schemes \cite{CCS2013, HealthSec2012, Vibration, RasmussenCCS2009, hei2014, NFC2} can be adopted to protect the IMD under such circumstances.

\subsection{Access Control Model}

There are three types of traditional access control models that may be applied in IMD context.

\begin{itemize}[leftmargin=*]
\item \textbf{Identity-based access control (IBAC):} a user is permitted or denied access based on whether the user appears in the access control list that contains all of the authorized users. 
\item \textbf{Role-based access control (RBAC):} the access permission is granted to a group of people who have a common role. 
\item \textbf{Attribute-based access control (ABAC):} a set of attributes are created and assigned to subjects to enforce the access control. The access rule is defined as a mixing of attributes, and the decision is made by matching the attributes required.
\end{itemize}

In fact, most existing access control schemes are not designed under traditional models, but instead are simplified to accommodate the resource-limited IMD:
\begin{itemize}[leftmargin=*]
\item \textbf{Pre-shared secret based access control:} Some works assume that the IMD and the programmer have pre-shared secret like a master key \cite{SP2008} or rolling code \cite{HealthCom2011}. However, considering that the IMD may be accessed by any doctor in emergency situations, the pre-distribution of the secret of each IMD to all possible doctors is not practical. 
\item \textbf{Proximity-based access control:} Some existing schemes manage access control solely based on physical proximity \cite{Hei2011, DenningCHI2010, CCS2013, HealthSec2012, Vibration, RasmussenCCS2009, hei2014, NFC2}. Although proximity is sufficient under the adversary model that the attacker will not approach the patient or make physical contact, it can only provide coarse-grained access control in which the identity of the requesting programmer operator is not authenticated.
\end{itemize}

To enable the authentication of the programmer operator, public-key cryptography must be used, which is feasible in the proxy-based access control architecture. However, considering that online authentication is unavailable and it is not practical to store the information of the huge number of eligible medical personnel into a local ACL on the proxy, IBAC model is not a viable option in IMD context. 
%
Instead, ABAC allows the multi-dimensional rules/policies (not just based on the identity or a simple role) to be enforced for fine-grained access control, e.g., the specialty and affiliation of the physician, the certificate of the eligibility to operate a certain model of IMD, etc.  

Therefore, our scheme employs the ABAC model to verify the qualifications of the access requestor. Meanwhile, to provide accountability in medical disputes, the proxy also authenticates the programmer operator and (if authorized) will record the access details (i.e., start time, end time) into a log as the evidence.

\subsection{Communications through Audio Cable}

\vspace{-1mm}
In the proxy-based access control architecture, the proxy needs to first build a secured connection with the IMD programmer. However, for general-purpose proxy device like smartphone, setting up a local connection with an external device (e.g. IMD programmer) via Bluetooth, NFC, or USB requires either the phone to be in the unlocked mode or the manual approval on the smartphone (also when the phone is unlocked). 
This means such types of connections will be unavailable when the smartphone is locked and the patient has gone unconscious, as the programmer operator does not know the password to unlock the patient's phone.
According to the recent data from Duo Labs, 34\% Android smartphones are not secured with a lock-screen passcode \cite{lock}. In another word, about 2/3 of phone users have enabled the screen-locking function. 
Therefore, we must choose a connection channel that is secure and available regardless of the patient's involvement.

We found that most modern smartphones have a headphone jack/port and most commercial IMD programmers have a USB port. A smartphone serving as the proxy can be connected with an IMD programmer through an audio cable. As shown in Figure~\ref{fig:wired}, the one end of the audio cable is plugged into the smartphone's headphone port while the other end links to an audio-to-USB adapter, which is then plugged into the USB port of the programmer. The audio-to-USB adapter is actually an external sound card with digital-to-analog converter (DAC) and analog-to-digital converter (ADC), hence the analog signal transmitting over the audio cable can be converted into digital data, or vice versa. The access control mobile application is always running in the background, ready to process the incoming requests from the audio cable.
The advantages of using the audio cable for communications include:
\begin{itemize}[leftmargin=*]
\item \textbf{No patient involvement required.} The data can be transmitted through audio cable in a plug-and-play manner, even if the phone is in the lock-up mode. 
\item \textbf{Reduce the attack surface.} The packets exchanged are no longer exposed in the air. The remote adversary cannot overhear or jam the communications. 
\item \textbf{Reduce the energy consumption.} Wireless transmissions consume at least 10 times more power than wired transmissions when providing comparable access rates and traffic volumes \cite{consumption}. Although both proxy and programmer are assumed to have sufficient power, energy saving becomes a critical concern when the phone battery is low.
\item \textbf{Proof of proximity.} The programmer operator needs to plug the audio cable into the headphone port of the smartphone, which proves the proximity of the operator to the patient. 
\item \textbf{Low cost.} The audio cable connection does not require extra hardware. An audio-to-USB adapter only costs around \$10.
\end{itemize}

\section{Attribute-based Encryption}

Our scheme achieves attribute-based access control using the attribute-based encryption (ABE) technique. ABE is a one-to-many encryption method, which allows data to be encrypted based on a set of attributes, so that only those users who own the specified attributes are able to correctly decrypt. ABE is collusion-resistant, which can prevent colluding users to gain access by combining their associated attributes, if none of them possesses the full set of required attributes.

There are two major types of ABE schemes: Key-Policy Attribute-Based Encryption (KP-ABE) \cite{KPABE} and Ciphertext-Policy Attribute-Based Encryption (CP-ABE) \cite{CPABE}. In KP-ABE, users' secret keys are generated based on an access tree (i.e. access policy) whose leaves are associated with attributes, and the data are encrypted over a set of attributes. Since the access policy is embedded in the decryptor's secret keys, the data encryptor has no control over who can access the data. However, in the IMD access control context, the proxy must check the programmer operator's qualifications, which are formed as a policy composed of a specific set of attributes. This requirement can be satisfied by CP-ABE, in which the users' secret keys are generated over a set of attributes and the ciphertext specifies the access policy.

Therefore, we adopt CP-ABE to implement the IMD access control. Specifically, the proxy encrypts the verification message with CP-ABE and sends the ciphertext to the programmer. The programmer can successfully decrypt the message only if the operator's attributes satisfy the access policy specified in the ciphertext. In contrast, the adversaries cannot decrypt the ciphertext, even if they collude. The CP-ABE scheme consists of the following four fundamental algorithms:
\begin{itemize}[leftmargin=*]
\item \textbf{Setup$(k)$}. The Setup algorithm takes a security parameter $k$ as input and randomly picks two exponents, to calculate the public parameters $PK$ and the master key $MK$. $PK$ will be used for encryption, while $MK$ will be used to generate users' secret keys and is known only to the central authority. 
\item \textbf{Encryption$(PK, M, T)$}. The encryption algorithm takes as input the public parameters $PK$, a plaintext message $M$, and an access tree structure $T$ over the universe of attributes. This algorithm will encrypt $M$, and produce a ciphertext $CT$ which only users who possess a set of attributes that satisfies the access structure $T$ are able to decrypt.  
\item \textbf{Key Generation$(MK, S)$}. The secret key generation algorithm takes as input the set of attributes $S$ that user $U$ owns, the master key $MK$, and randomly selects a set of $|S|+1$ numbers specific to user $U$. It outputs a secret key $SK$. 
\item \textbf{Decrypt$(PK, CT, SK)$}. The decryption algorithm takes as input the public parameters $PK$, a ciphertext $CT$ which contains the access policy $T$, and a secret key $SK$ generated from attribute set $S$. If the set $S$ of attributes satisfies the access policy $T$, the algorithm will successfully decrypt the ciphertext and return the plaintext message $M$.
\end{itemize}

\begin{figure*}[t]
\centering
\includegraphics[width=6in,height=2.8in]{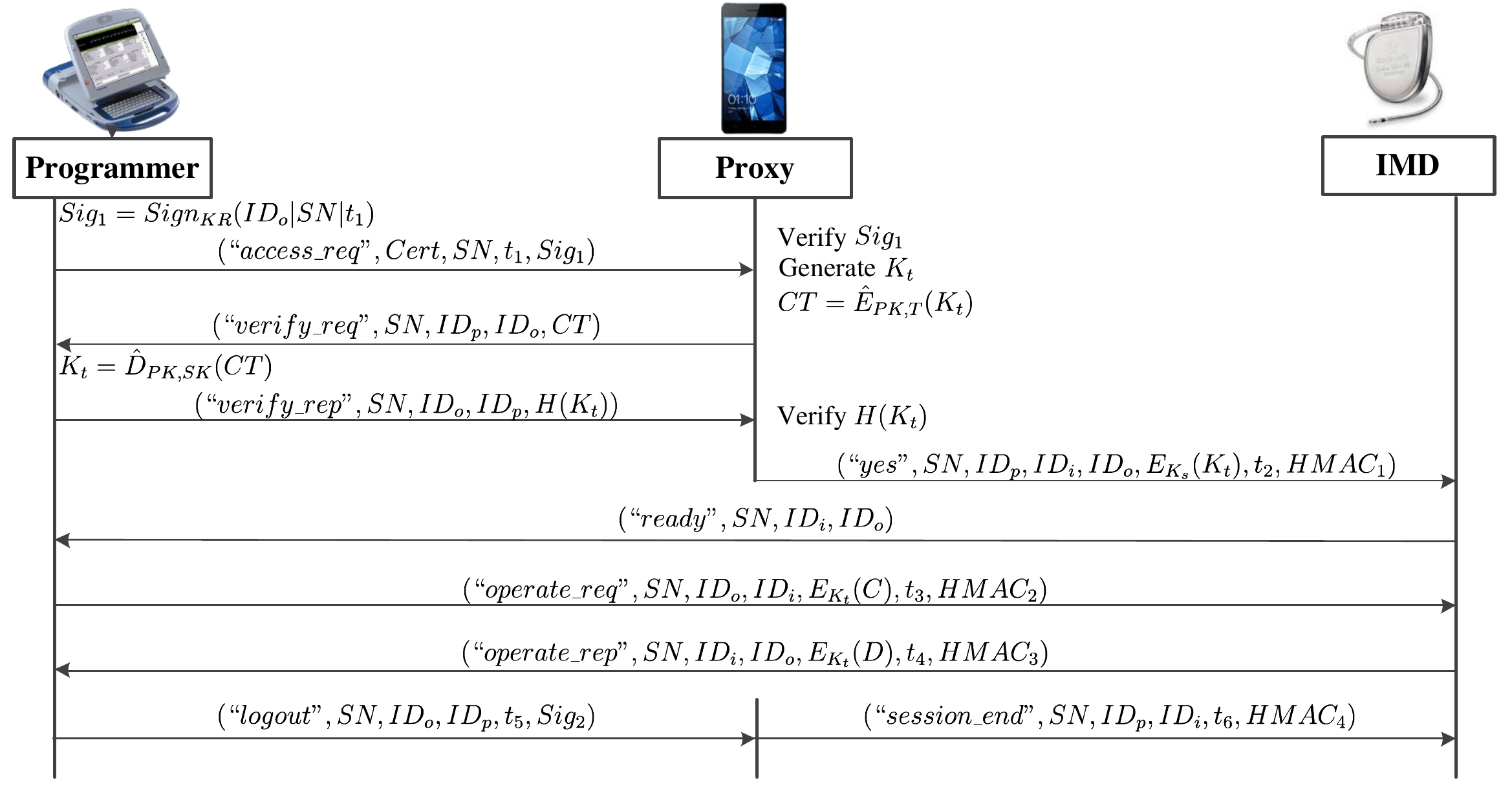}
\vspace{-2.5mm}
\caption{Access Control Procedure}
\vspace{-5mm}
\label{fig:workflow}
\end{figure*}

\section{Protocol Design}


\subsection{System Overview}
\begin{itemize}[leftmargin=*]
\item \textbf{IMD.} Each IMD has a unique identification $ID_{i}$ and a master key $K^{M}_{i}$. Note that the master key is only used for pairing up the IMD with a proxy device, and will not directly participate the access control procedure.
\item \textbf{Programmer.} The programmer can be simply viewed as the terminal device used by its operator to interact with the IMD or proxy. It obtains all information required for access control (e.g., secret keys, certificates) from the operator, by manual input or reading in from a smart card.
\item \textbf{Operator.} The programmer operator is the actual subject to be verified. All legitimate operators must first be registered at a Central Health Authority (CHA), which manages the qualifications of operators and issues digital certificates for them. Each operator has a unique identification $ID_{o}$, a pair of public/private keys $KU$ and $KR$, and a public key certificate $Cert$. The qualifications that an operator owns correspond to a set of attributes $S$. CHA will generate the secret key $SK$ for the operator based on the set of attributes $S$. $SK$ can be used to decrypt the ciphertext produced by CP-ABE if the access policy is satisfied. Besides, all operators know the public parameters $PK$ used in CP-ABE.   
\item \textbf{Proxy.} The proxy device has the identification $ID_{p}$. There is a client program running on the proxy to perform the access control for the IMD. The proxy has been paired up with the IMD through initial setup. 
The client program has a copy of the public parameters $PK$ used to run CP-ABE, and is able to generate the access tree (policy) $T$ that describes the qualifications required for access.
\end{itemize}


%

\subsection{Access Control Protocol Design}

The access control procedure includes two separate processes: the authentication of the programmer operator and the authorization for access. 

Although the IBAC model is not suitable in IMD context and we utilize attributes to control access instead, it is still necessary to authenticate the identity of the programmer operator. One reason for that is to provide non-repudiation guarantee in case of medical disputes. For example, a programmer operator cannot deny his/her access if the start time and end time of the access have been signed by his/her own private key. The other reason is that the ciphertext $CT$ generated by CP-ABE contains the access policy in plaintext. The access policy specifies the expected qualifications of the authorized physicians (e.g., speciality) and the information related to the IMD model (e.g., the certifications required to operate), which are all very sensitive with regard to the patient's privacy and should not be publicly accessible to anyone who requests for access. Therefore, our scheme authenticates the programmer operator before the authorization stage, so that only a legitimate operator who has registered at CHA (not necessarily authorized to access) can continue to the authorization process and view the access policy.

In the authorization stage, the proxy encrypts a randomly generated temporary session key $K_{t}$ with CP-ABE and sends the ciphertext to the programmer. If the operator is an eligible physician whose qualifications (attributes) satisfy the access policy, the session key $K_{t}$ can be correctly retrieved and used to establish a secured communication channel with the IMD.

The access control procedure is presented in Figure~\ref{fig:workflow}. 
We assume that the proxy device has already been paired up with the IMD, and a pair of symmetric keys $K_{s}$ have been shared between them for encrypted communications. 
The detailed procedure is described as follows:
\begin{enumerate}[leftmargin=*]
\item The programmer initializes the access control protocol by connecting with the proxy via audio cable, and sending an access request which is composed of a unique action sequence ``$access\_req$'', the operator's digital certificate $Cert$, a random selected session number $SN$, timestamp $t_{1}$, and a signature $Sig_{1}$ signed by the operator's private key $KR$. The certificate contains the operator's public key $KP$ and identification $ID_{o}$. The signature $Sig_{1}=Sign_{KR}(ID_{o}|SN|t_{1})$ is attached to prove that the current access requestor is indeed $ID_{o}$. 
\item The access control mobile application has registered a receiver of the headset connection state changes. When this program is notified of the plug-in event, it will read in and demodulate the audio data. If the action sequence ``$access\_req$'' can be found in the demodulated data, it indicates that the data is for IMD access request instead of regular audio (e.g., music). Then it extracts and verifies the received $Sig_{1}$ using the requestor's certified public key $KP$ embedded in $Cert$. If the signature is valid, the programmer operator is successfully authenticated. The proxy will next check if that operator is authorized to access. Specifically, it randomly generates a temporary session key $K_{t}$ and encrypts $K_{t}$ using CP-ABE. Then, the produced ciphertext $CT = \hat{E}_{PK, T}(K_{t})$ is sent back to the programmer. 
\item The programmer decrypts $CT$ with the operator's secret key $SK$ (generated and assigned by the CHA). If the operator's qualifications (attributes) satisfy the access policy, the secret key $SK$ will be able to retrieve the temporary session key $K_{t}$. Then, it calculates the hash value of $K_{t}$ and sends the hash value to the proxy.
\item The proxy also calculates the hash value of $K_{t}$ with the same hash function. If the two hash values are equal, it indicates that the programmer operator is eligible for access. The proxy will inform the IMD that the programmer $ID_{o}$ has been authenticated and is authorized to access, and sends a copy of the session key $K_{t}$ to IMD. Note that all communications to/from the IMD are conducted in the wireless channel, which may suffer eavesdropping, replay, and tampering attacks. Hence, the session key is encrypted by the shared key $K_{s}$ to prevent eavesdropping. A timestamp $t_{2}$ is added to defend replay attacks. The keyed-hash message authentication code (HMAC) of the message is calculated using $K_{s}$ to ensure the authenticity and integrity of the message. 
\item After receiving the authorization notification from the proxy, the IMD retrieves the $K_{t}$ and sends ``ready'' message to the programmer. 
\item In the mutual communications between the IMD and the programmer, the operation commands $C$ sent by the programmer and the data/result $D$ returned by the IMD are all encrypted using the temporary session key $K_{t}$. Each authorization permits multiple operations (e.g., data reading, drug delivery). We only draw one round of operation in Figure~\ref{fig:workflow} for illustration. The timestamps and HMACs are also adopted in the communications between the IMD and the programmer to defeat various active wireless attacks.
\item After the programmer has completed the access, it sends a ``logout'' notification message to the proxy. Another signature $Sig_{2}=Sign_{KR}(ID_{o}|SN|t_{5})$ is generated in which the access end time $t_{5}$ is signed. 
\item Finally, the proxy will notify the IMD that the current session has ended so that the session key $K_{t}$ will be removed. Timestamp $t_{6}$ and $HMAC_{4}$ are included in the this message.  
\end{enumerate}

Our scheme asks the programmer to explicitly log out the session, and requires it to sign the time that session ends. Therefore, $Sig_{1}$ and $Sig_{2}$ together can prove that the operator has accessed the IMD in that period of time. Any wrong operations performed in this period will be attributed to that specific operator. The programmer should maintain the wired connection with the proxy during interacting with the IMD, to eliminate the possibility that a second programmer is connected with the proxy while the first one is still interacting with the IMD. 
However, it may happen that the programmer does not sign out by the end of its session for certain reasons, a time-out mechanism is used to tackle this situation. Specifically, if there is no interaction made by the programmer for a fixed amount of time $T_{out}$, the session will be closed and the session key $K_{t}$ will be disabled.

%



\begin{figure}[t]
\centering
\includegraphics[width=3.4in,height=1.8in]{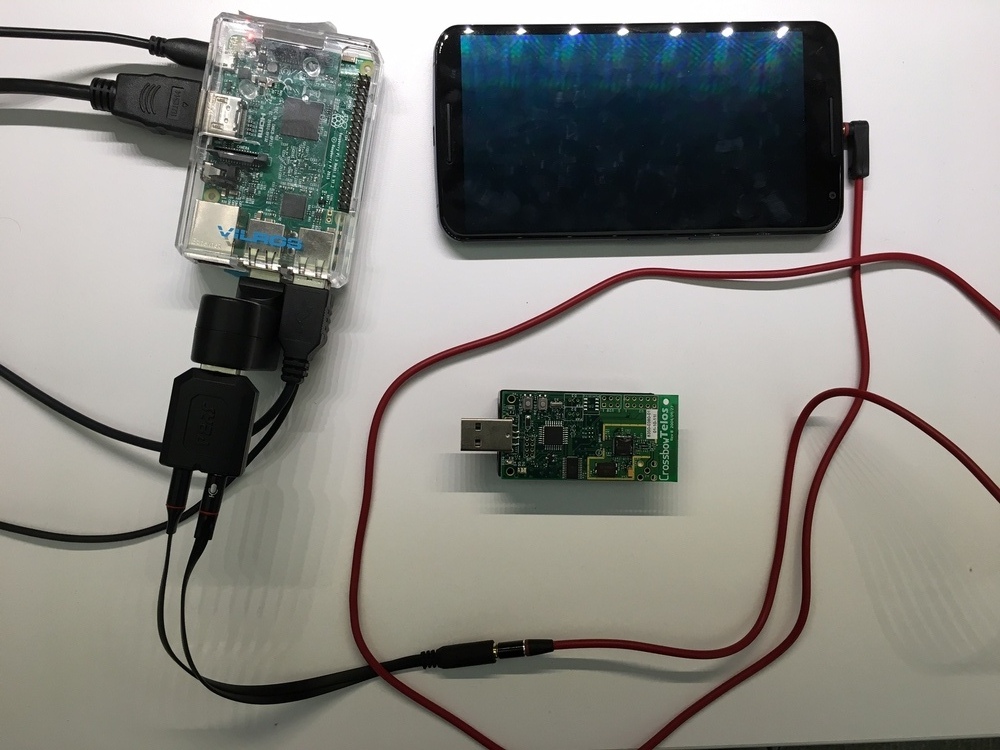}
\caption{Prototype Setup}
\vspace{-1mm}
\label{fig:implement}
\vspace{-1mm}
\end{figure}

\begin{figure}[t]
\centering
\includegraphics[width=3.45in,height=1.8in]{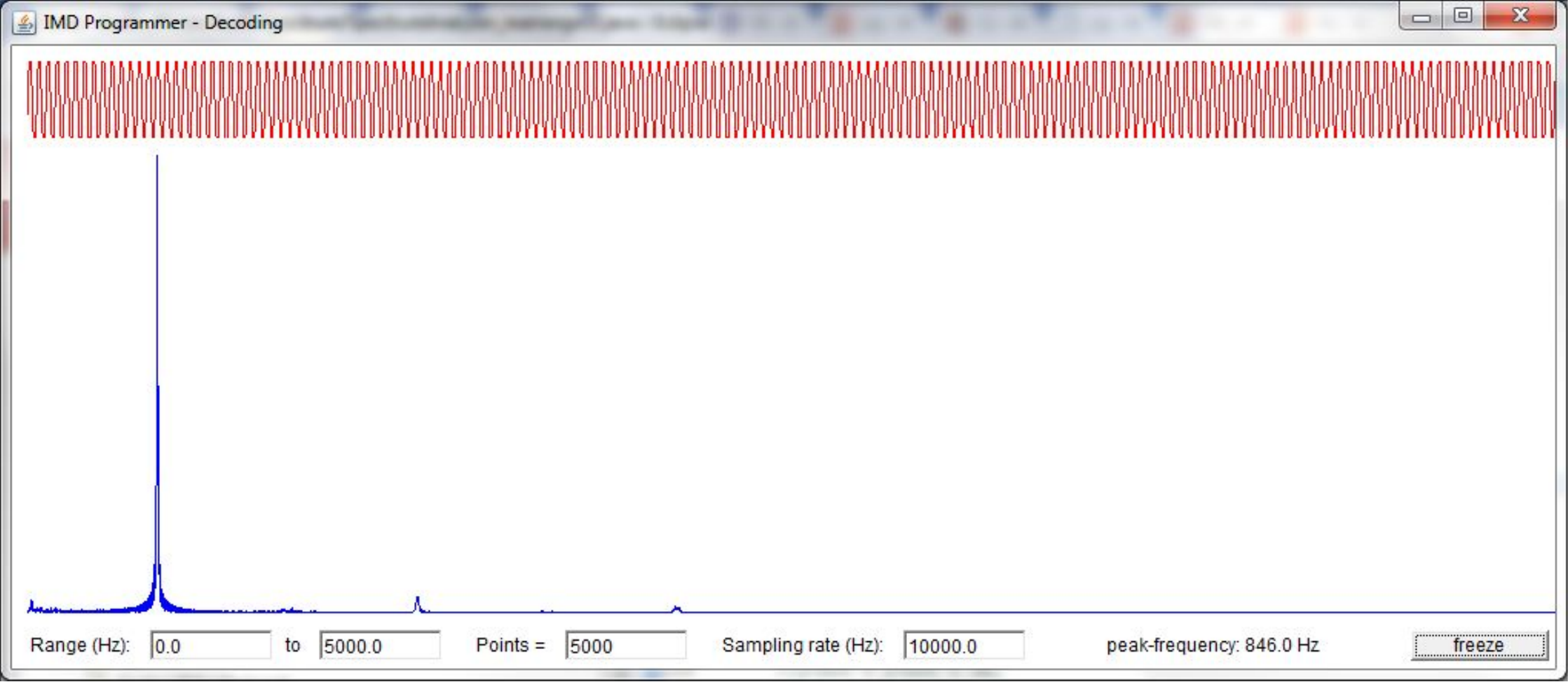}
\caption{Illustration of Signal Modulated by BFSK}
\label{fig:dft}
\vspace{-1mm}
\vspace{-4mm}
\end{figure}

\section{Security Analysis}

\vspace{-1.5 mm}
\subsection{Resistance to Passive Attacks}

In the wireless channel, the adversary is able to eavesdrop the communications between the proxy and the IMD, as well as between the programmer and the IMD, to obtain the data, operation commands, or keys being transmitted. To defend against such passive wireless attack, all sensitive information is protected with symmetric encryptions. Specifically, a pair of keys $K_{s}$ have been pre-shared by IMD and the proxy (during initial setup), and another pair of keys $K_{t}$ are securely distributed to IMD and the programmer. The adversary cannot decrypt the ciphertexts without the cryptographic keys.

Over the audio cable, even if the proxy has been infected by malware which will overhear the audio data received by the proxy, our scheme can ensure that it cannot obtain any sensitive information. The temporary session key $K_{t}$ is encrypted by CP-ABE. The malware cannot correctly derive the plaintexts without the required qualifications (attributes). Additionally, it cannot infer $K_{t}$ from its hash value $H(K_{t})$.


\subsection{Resistance to Active Attacks}

In the wireless channel, all messages containing sensitive information are timestamped to prevent replay attacks. In addition, a HMAC is calculated using the corresponding shared keys ($K_{s}$ between the proxy and the IMD, or $K_{t}$ between the programmer and the IMD), to defeat the tampering attacks.

Over the audio cable, the malware on the proxy can send arbitrary messages to the programmer or tamper the messages sent by the application running the access control. In our scheme, the only outgoing message to the programmer via the wired audio channel is $CT$. However, either the replay or manipulation of this message cannot help an unauthorized programmer operator to gain access, since only eligible operators can successfully decrypt $CT$ and retrieve the session key $K_{t}$.

\subsection{Other Attacks}

Over the audio cable, the mobile malware can add noise to the audio data sent by the access control application, but this sort of attacks can only disrupt the access of legitimate operators (e.g. causing the failure of authentication or authorization). As we have mentioned before, DoS attacks are not the focus of this work.
Besides, our scheme is well-compatible with existing solutions to the proxy-absence spoofing attacks \cite{IMDGuard} and the resource depletion attacks \cite{Liu,SIGCOMM,Hei2010}.



\section{Evaluation}

In this section, we implement our schemes on real devices and evaluate the overheads of our scheme with experiments.

\subsection{Prototype Implementation}

The key challenge of the implementation of our scheme is the difficulty in obtaining open-source commercial IMD products. Alternatively, in our prototype system, we use TelosB Mote TPR2420 sensor node with 8 MHz TI MSP430 microcontroller, 10kB RAM, and 48kB Flash Memory as the replacement of IMD. We choose the Rasberry Pi Version 3 Model B, a small single-board device with 1.2GHz 64-bit quad-core CPU and 1GB RAM, to emulate the programmer. A Nexus 4 smartphone powered by a 1.5 GHz quad-core processor with 2 GB of RAM, is used to emulate the proxy.

In our scheme, symmetric encryption and public-key cryptography are implemented using the Advanced Encryption Standard (AES) algorithm and RSA algorithm, respectively. We use 128 bits key for AES encryption and 1024 bit keys for RSA encryption. SHA-1 is chosen for HMAC generation.


\begin{table}[t]
  \caption{Parameters used for modulation and demodulation}
  \footnotesize
  \centering
  \begin{tabular}{|m{1.3in}<{\centering}|m{1in}<{\centering}|}
    \hline
    Parameter    &   Value  \\
    \hline
    Sampling rate   &  44100Hz  \\
    \hline
    Pulse-code modulation (PCM) bit depth   &  16 \\
    \hline
    $f_{0}$   &  1575Hz  \\
    \hline
    $f_{1}$   & 3150Hz\\
    \hline
    Baud rate   &  315 \\
    \hline
  \end{tabular}
  \label{table:modu}
\vspace{-3.5 mm}
\end{table}

\subsection{Testbed Specification}

We developed an Android application to delegate the access control on the Nexus 4 smartphone. This mobile application is able to perform CP-ABE and modulate/demodulate the audio signals. Correspondingly, we developed a client program running on the Raspberry Pi, which can decrypt the ciphertext encrypted using CP-ABE if the programmer operator's attributes satisfy the access policy embedded in the ciphertext. This client program is also capable of modulating/demodulating audio signals. 
The smartphone and the programmer is connected via an audio cable and a SYBA external USB Sound Adapter (audio-to-USB convertor). A screenshot of the prototype is shown in Figure~\ref{fig:implement}.

We adopted the Binary Frequency-Shift Keying (BFSK) frequency modulation scheme for modulation, in which the digital data are converted into the analog signals at two different frequencies for transmissions over the audio cable. For example, the binary ``0'' bit is represented by the audio signal at frequency $f_{0}$ while the binary ``1'' bit is represented by the audio signal at frequency $f_{1}$.  Figure~\ref{fig:dft} displays the results of the signal modulation and spectrum analysis.
The red signal (upper sine waves) in Figure~\ref{fig:dft} is the analog signal after modulated by BFSK. 
At the receiver end, the analog signal is sampled to a sequence of discrete-time signal (samples). Then, we use the Discrete Fourier Transform (DFT) algorithm to convert the sampled analog signal from time domain into the frequency domain representation, which is illustrated in Figure~\ref{fig:dft} as the blue pulse-like signal. Specifically, the analog signal to be demodulated can be viewed as an addition of multiple sine signals in different frequencies. With Fourier transform, the magnitudes of the modulated signal on various frequencies within the spectrum range are calculated.
The frequency with the highest amplitude (i.e., maximum power) is called the peak frequency. If the peak frequency equals $f_{0}$, then the current signal sample represents a ``0'' bit; while if the peak frequency equals $f_{0}$, then the signal sample represents a ``1'' bit.
The parameters we used for modulation and demodulation are listed in Table~\ref{table:modu}:


%


\begin{figure}[t]
\centering
\begin{minipage}{1.65in}
  \centering
  \includegraphics[width=1.65in,height=1.55in]{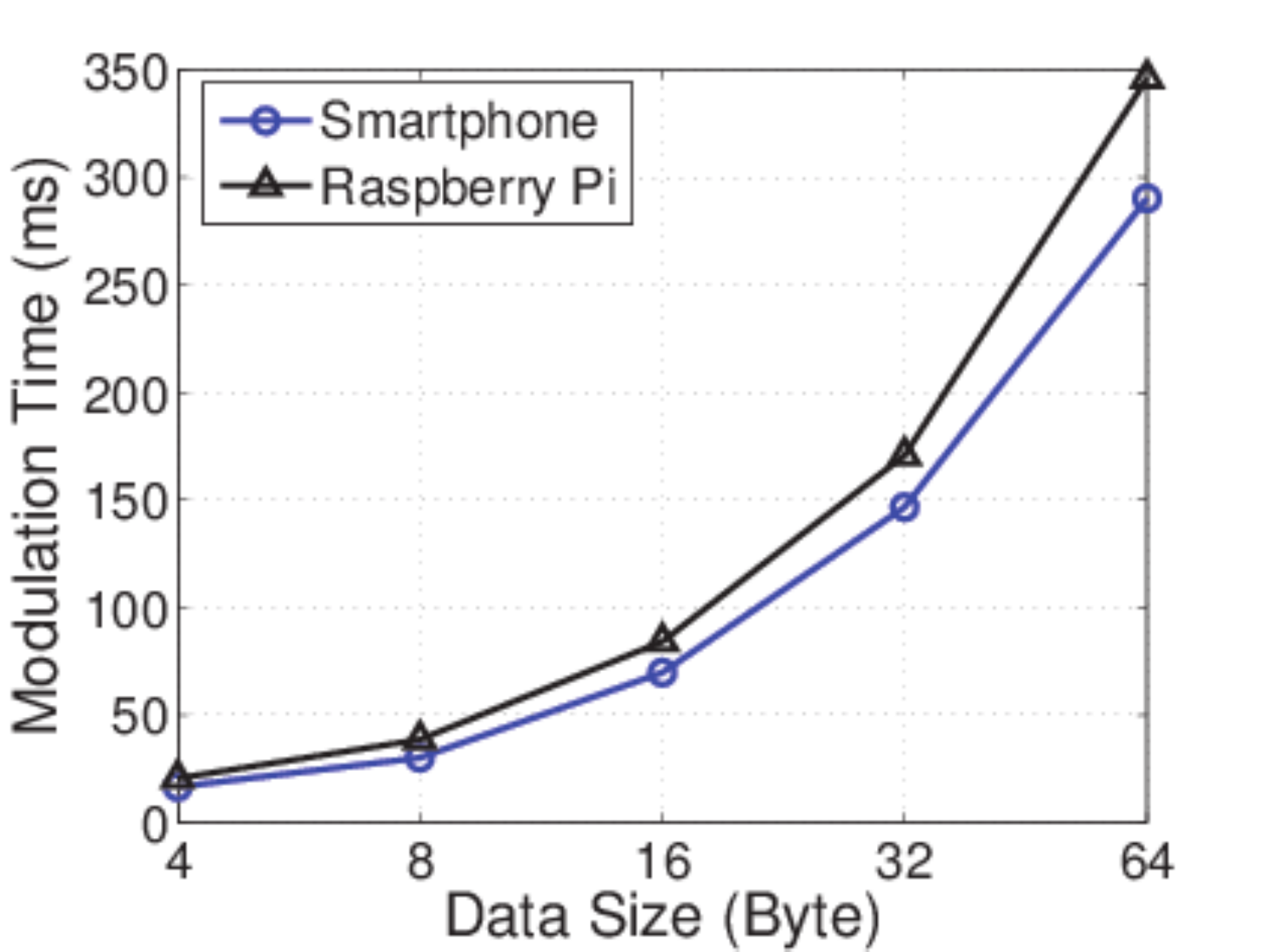}
	\vspace{-7mm}
  \caption{Time - Modulation}
  \label{fig:modulation}
\end{minipage}%
\hspace{2mm}
\begin{minipage}{1.65in}
  \centering
  \includegraphics[width=1.65in,height=1.55in]{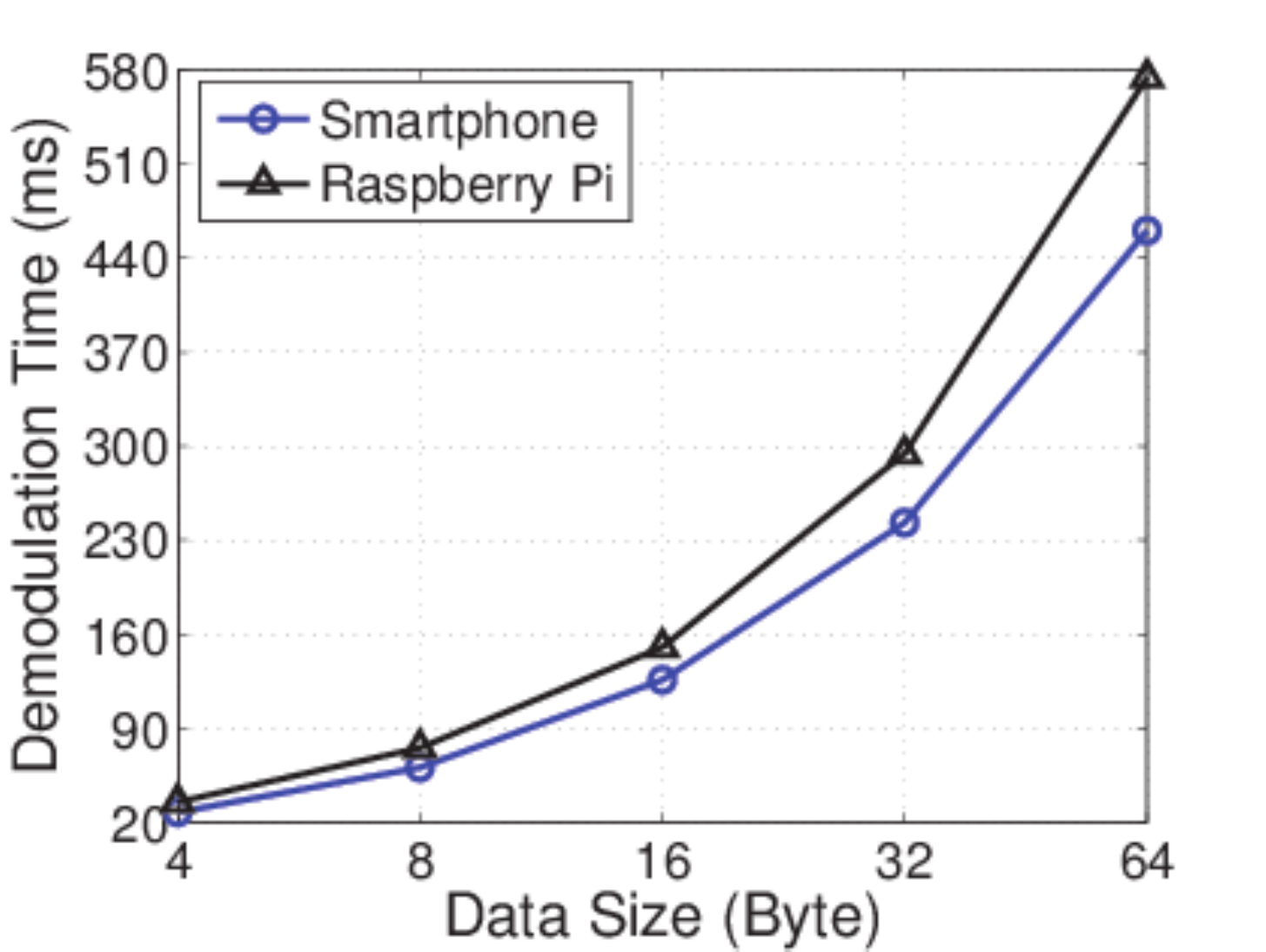}
	\vspace{-7mm}
  \caption{Time - Demodulation}
  \label{fig:demodulation}
\end{minipage}
\vspace{-2.5 mm}
\end{figure}


 
%

\begin{figure}[t]
\centering
\begin{minipage}{1.65in}
  \centering
  \includegraphics[width=1.65in,height=1.5in]{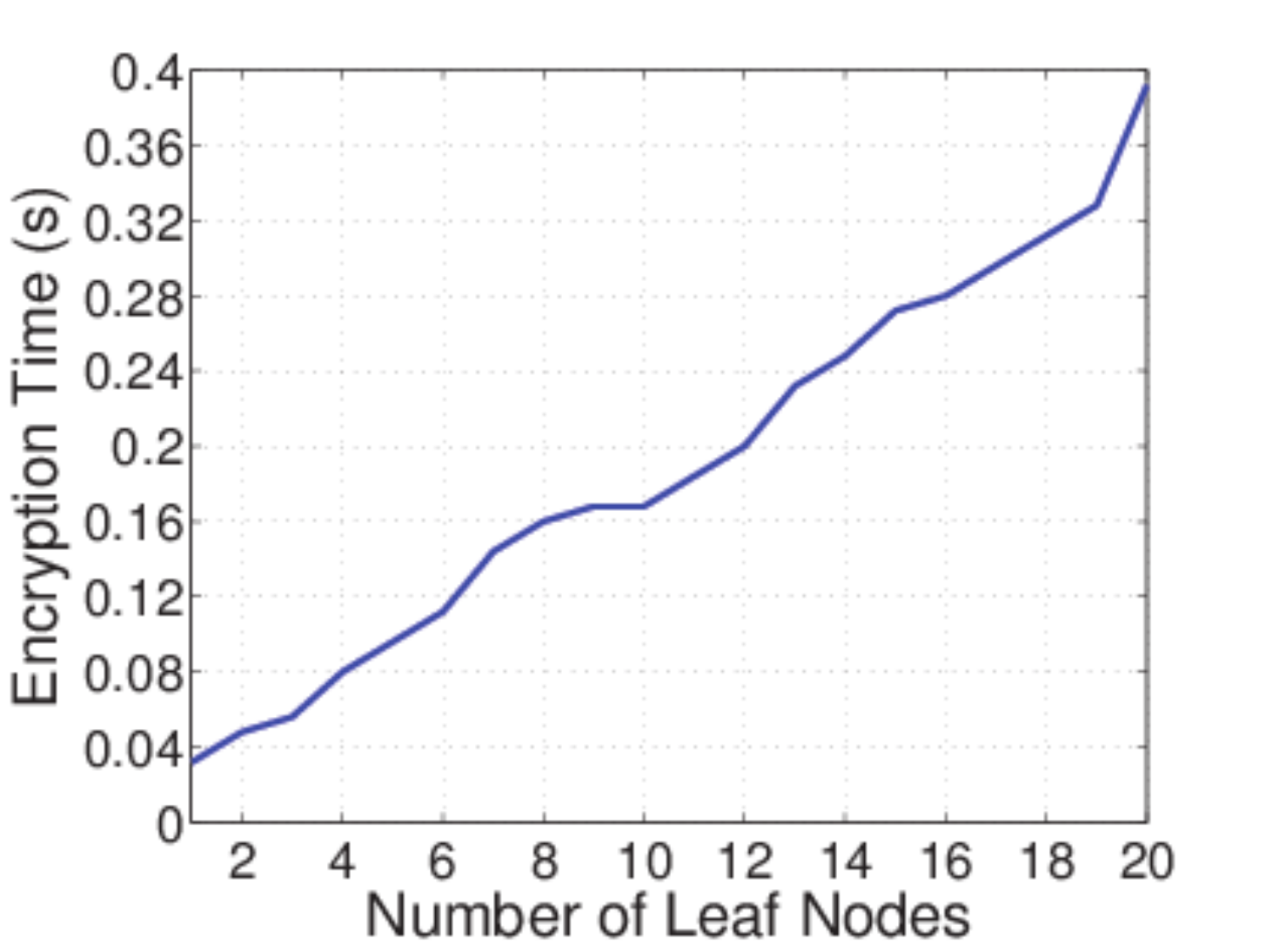}
	\vspace{-6.5mm}
  \caption{Time - CP-ABE Encryption}
  \label{fig:E}
\end{minipage}%
\hspace{2mm}
\begin{minipage}{1.65in}
  \centering
  \includegraphics[width=1.65in,height=1.5in]{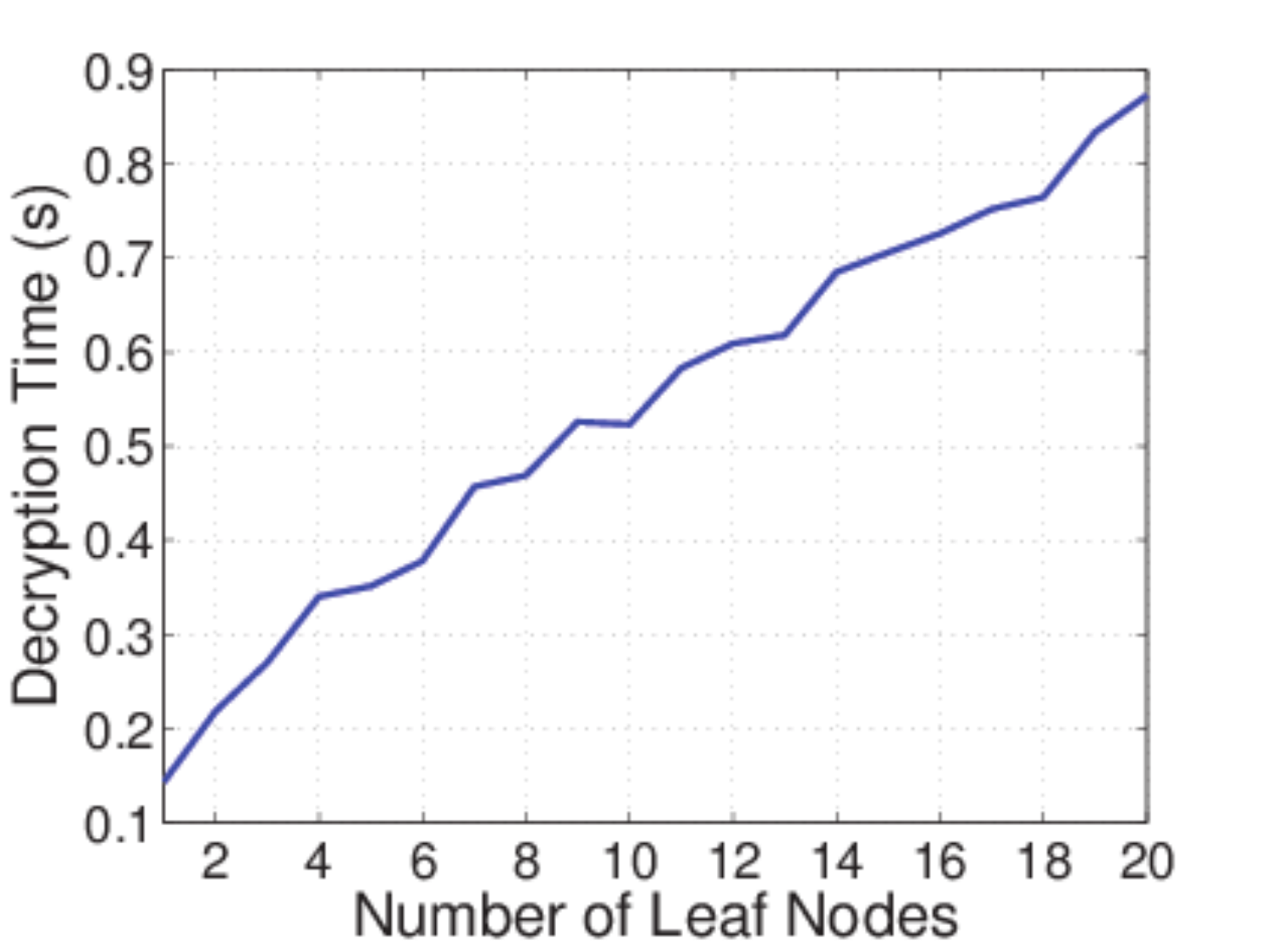}
	\vspace{-6.5mm}
  \caption{Time - CP-ABE Decryption}
  \label{fig:D}
\end{minipage}
\vspace{-4.5 mm}
\end{figure}

\subsection{Experimental Results}

To evaluate the efficiency of our scheme, we measure the computational overheads of the protocols running on the IMD (TelosB Mote), the proxy (Nexus 4 smartphone), and the  IMD programmer (Rasberry Pi), respectively. All the run-time overheads are the average of 50 measurements.

\subsubsection{IMD} The major cryptographic computations performed on the IMD are the symmetric encryption and HMAC. On TelosB node, the 128-bit AES encryption takes 2ms. For HMAC, we estimate the length of the plaintext message $(``operate\_rep\textquotedblright, SN, ID_{i}, ID_{o}, E_{K_{t}(D)}, t_{4})$ to be 78 bytes in total, including a 4-byte command, a 2-byte sequence number, two 4-byte IDs, and a 64-byte data/result returned). The HMAC computation over a 78-byte massage takes 47ms.

\subsubsection{IMD programmer and proxy} The IMD programmer (Rasberry Pi) and proxy (smartphone) both need to conduct modulation/demodulation for their wired communications through the audio cable. Figure~\ref{fig:modulation} and Figure ~\ref{fig:demodulation} show the time consumption for modulation and demodulation on smartphone and Rasberry Pi, respectively. As we can see, the Rasberry Pi has a better performance than the smartphone, and the demodulation takes longer times than the modulation process.

Additionally, the proxy (smartphone) needs to encrypt the temporary session key with CP-ABE encryption algorithm, and he IMD programmer (Rasberry Pi) will run the decryption algorithm to retrieve the key. Figure~\ref{fig:E} and Figure~\ref{fig:D} show the time consumption for CP-ABE encryption on smartphone and decryption on Rasberry Pi, respectively. The run-time overheads for CP-ABE encryption and decryption both increase with the number of leaf nodes (attributes). Our experiments tested a maximum of 20 attributes, which should be sufficient to specify the qualifications of the programmer operator. The decryption is found to be the most time-consuming step in the whole scheme, which takes about twice the time used for encryption. Since the CP-ABE encryption/decryption is required only once, their execution time are considered acceptable.


%
%
%
%
%
%
%
%
%
%
%
%
%
%
%
%
%
%
%
%

\section{Related Work}

Most existing works employ the two-party access control architecture, in which the IMD with constrained computational power, battery capacity and storage must perform the access control by itself, resulting in a relatively weak access control and shorter IMD lifetime. Some works proposed that the IMD should be accessible only to a group of trusted people (e.g., doctors, relatives) who have been added into the IMD's access control list (ACL) \cite{ACL1}\cite{ACL2}. The limited storage of IMD will greatly restrict the scalability of such schemes. In addition, the authentication of the requestor requires the verification of the certificate/signature. The computing power and battery can hardly afford the public-key cryptographic computations.

Some other two-party based access control schemes only check whether the programmer is in proximity. Specifically, IMD and the programmer need to extract certain features from the same source simultaneously, and generate the temporary keys based on the extracted feature, respectively. The source is usually the signal in an out-of-band channel, such as electrocardiography (ECG) signal \cite{CCS2013}, body-coupled electric signal \cite{HealthSec2012}, vibration \cite{Vibration}, ultrasound \cite{RasmussenCCS2009}, etc. If the programmer is close enough to the IMD, they will derive the same (symmetric) temporary key for the encryption of future communications. The real-time signal measurement and key extraction computations both bring in an extra burden to the resource-constrained IMD.
Other works proposed to pre-load the patient's biometric information (e.g., fingerprint) or a password into the IMD. During access, the programmer operator can manually collect the biometric features from the patient \cite{Hei2011} or the password from a physical object carrying it \cite{DenningCHI2010}, respectively. However, the bio-features and the password engraved on an object can be stole by the adversary. Besides, a common disadvantage of these two types of schemes is that they only provide a coarse-grained access control. No information about the access requestor is acquired and validated.

By contrast, in the proxy-based architecture, the powerful proxy device handles those resource-consuming tasks. It can support more complicated and robust access control schemes. However, previous proxy-based access control schemes either depend on anomaly detection of the access pattern \cite{Hei2010,MedMon} or only verify the authenticity of the programmer \cite{IMDGuard,Cloaker}. The former method does not check the access requestor and require a training process for each patient. The second method is vulnerable considering that an attacker could also purchase or steal a legitimate programmer. Our scheme views the programmer and its operator together as one subject, and the access decision is made based on the programmer operator. We not only authenticate the operator, but also employ the CP-ABE algorithm to verify the qualifications of the operator.

Key management \cite{management1, management2, management3} is essential for security. Several papers \cite{Hei2011,other2,GuanIoTJ,other3,other4,GuanCommMaga,other5} have studied related security issues.

\section{Conclusion}

%

In this paper, we proposed a novel fine-grained IMD access control scheme based on a proxy device like the patient's smartphone, which will delegate the heavy access control computations for the IMDs. To mitigate the potential wireless attacks, the communications between the proxy and the IMD programmer are conducted through an audio cable. The ciphertext-policy attribute-based encryption is employed to enforce the fine-grained access control over the qualifications of the programmer operator. We built a prototype to evaluate our scheme. The experimental results demonstrated its feasibility and effectiveness.

\bibliographystyle{abbrv}
\bibliography{./IEEEabrv,./IEEEexample}

\end{document}